\documentclass[preprint]{aastex701}

\usepackage{graphicx}%
\usepackage{multirow}%
\usepackage{amsmath,amssymb,amsfonts}%
\usepackage[title]{appendix}%
\usepackage{xcolor}%
\usepackage{soul}
\usepackage{hyperref}
\usepackage[export]{adjustbox}

\begin{document}

\title{An Acoustic Model for Sunquakes Unifying the Solar Interior and Atmosphere}
\correspondingauthor{John T. Stefan}
\author[orcid=0000-0002-5519-8291,gname='John T.',sname=Stefan]{John T. Stefan}
\affiliation{New Jersey Institute of Technology, Department of Physics}
\email[show]{john.stefan@njit.edu}  

\author[orcid=0000-0003-0364-4883,gname='Alexander G.',sname=Kosovichev]{Alexander G. Kosovichev}
\affiliation{New Jersey Institute of Technology, Department of Physics}
\email{alexander.g.kosovichev@njit.edu}

\begin{abstract}

 One of the leading hypotheses for sunquake generation suggests that flare-accelerated particles originating from the reconnection site in the corona travel down to the chromosphere and photosphere, where they deposit energy through collisions and subsequently drive acoustic oscillations. To properly encompass this top-down excitation mechanism, we extend the domain of a semi-spectral 3D acoustic model of the global Sun up to several 10's of Mm above the photosphere, where the transition region and lower corona are resolved. We then use the radially-dependent heating rates derived from the flare radiative hydrodynamic (RADYN) simulations---extrapolated to a 3D profile---to realistically excite sunquakes. In addition to the usual sunquake wavefronts, we also observe waves that propagate through the chromosphere and corona in a similar fashion to Moreton-Ramsey waves and large-scale coronal propagating fronts (LCPFs). We examine the dynamics of these waves and discuss how they may be used to constrain models of sunquake excitation.

\end{abstract}


\section{Introduction}\label{sec:intro}
Sunquakes are impulsive helioseismic events observed on the solar surface in images of the photospheric line-of-sight velocity \citep{Kosovichev1998}. The energy released by solar flares is understood to be driven by magnetic reconnection \citep{Anti_reconn,Kopp_reconn}, the process that describes the breaking and reforming of magnetic field lines, which can also accelerate particles such as protons and electrons to very high energies. Such beams have been suggested to be a possible means of exciting sunquakes \citep{Sharykin_SQ,Macrae_SQ,Pedram_SQ}. In the so-called thick target model, electron beams strike the chromosphere, where the deposited energy results in a sharp increase in hard X-ray emission and may also generate downward-propagating shocks \citep{Fisher_TT}. Prior work studying the origin of sunquakes has identified the lowest layers of the Sun's atmosphere as likely locations of sunquake excitation \citep{KosZhar1995,Zharkova,Sharykin_SQ,Riuzhu}, and the heating and shocks associated with the energy deposited by the particle beams have been determined to be sufficient to excite helioseismic waves \citep{Kosovichev2015,KosZhar1995}. Additional attempts to explain the excitation of sunquakes included the impulsive magnetic force associated with CME eruptions \citep{Fisher2012}. However, a statistical study showed a strong correlation of sunquake sources with the rate of X-ray emission, indicating their association with particle beams \citep{Sharykin2020}. Recent observational and theoretical studies of sunquake photospheric sources suggest that direct heating and momentum transport by proton beams in the lower atmosphere during the flare impulsive phase are likely mechanisms of sunquakes \citep{Sadykov2023,Kosovichev2023}.

Given the expectation that the energy required to excite sunquakes originates in the solar atmosphere, it would seem natural to simulate sunquake generation with a domain and source within this region. A majority of past sunquake modeling efforts, however, have been limited to the solar interior or just slightly above it. For example, \citet{Podesta2005} employed a normal mode approach with a domain that terminates at the photosphere, with the remaining atmosphere and corresponding physics remaining unresolved. Earlier work performed by \citet{Medrek2000} included the chromosphere using a two-layer solar model with simplified stratification. Here, the authors directly simulate the non-linear, magnetic-free Euler equations for more detailed solutions, though the acoustic source remains submerged beneath the photosphere. More recently, non-linear MHD simulations using the Lare2d code \citep{Arber2001} that include the solar atmosphere up through the corona have shown that restructuring of the coronal magnetic field is capable of exciting downward-propagating acoustic waves \citep{Russell2016}. The domain for these simulations is relatively small, particularly in the horizontal direction, so the global propagation of the produced acoustic wave was not studied. We aim to address the deficiencies of these past models in capturing the excitation of a globally propagating sunquake from coronal sources by using a semi-spectral, linear acoustic model that incorporates the full solar interior coupled with realistic atmospheric stratification.

Our inclusion of the chromosphere and corona necessarily means that other wave-like phenomena that propagate in these regions should also be considered. Two particularly prominent phenomena are Moreton-Ramsey waves (hereafter referred to as Moreton waves) and large-scale coronal propagating fronts (LCPFs). The discovery of Moreton waves in the early '60s \citep{Moreton1960,Moreton1960a} was quickly followed by attempts to explain their apparent supersonic propagation speed. These waves are fast-propagating and relatively weak amplitude disturbances observed traveling through the chromosphere. One prominent model incorporated an initial propagation of a fast magnetosonic wave through the corona \citep{Uchida}, which in turn generates the Moreton wave as it interacts with the chromosphere. The subsequent discovery of a large-scale coronal wave \citep{Thompson}, the so-called EIT (Extreme Ultraviolet Imaging Telescope) wave, seemed to support this model; however, the propagation speed of this wave differed from the model’s prediction \citep{Klassen}.

This discrepancy can be explained given EIT's long observational cadence \citep{Byrne2013}, and evidence for the Moreton wave's coronal counterpart has been definitively identified in Extreme Ultraviolet (EUV) images from the Atmospheric Imaging Assembly (AIA) instrument aboard the Solar Dynamics Observatory \citep{Nitta,Cabezas}. These counterparts are generally referred to as global EUV waves, or more generally as large-scale coronal propagating fronts, and modeling efforts have reproduced their excitation with a simulated Coronal Mass Ejection (CME) \citep{Downs}. In such models, the expanding coronal loops of the CME compress the coronal plasma, and the compression front eventually decouples from the CME as a global wave that continues to propagate at the fast magnetosonic speed. More recent work has attempted to explain the lack of companion Moreton waves in EUV wave events by suggesting that inclined CMEs---which are relatively rare---are responsible for the chromospheric waves \citep{inclined1,inclined2}. While past analyses of global EUV waves have indicated that a vast majority ($~$95\%) have been excited in conjunction with a CME \citep{Muhr}, the generation of the remaining global EUV wave events cannot be explained by a CME origin.

In this work, we perform simulations of energy deposition into the solar atmosphere by incident proton and electron beams. We use the extended 1D radiative hydrodynamics RADYN code \citep{RADYN} to simulate the initial reaction of the solar atmosphere, and the resulting parameters are supplied to a 3D acoustic model to study wave generation and propagation \citep{SQ_model}. We not only confirm that these beams are capable of eliciting a helioseismic response, but that they may also simultaneously generate waves analogous to global EUV and Moreton waves. We note that as a numerical compromise for the inclusion of the chromospheric and coronal layers, we do not include magnetic effects that have been shown to be important in fully understanding the dynamics of atmospheric waves in low plasma beta environments \citep{Pian2018,Vrsnak2016}. Therefore, analogies with the chromospheric and coronal waves are purely qualitative, and we discuss how inclusion of the magnetic field might influence our results in Section \ref{sec:diss}. Still, this link provides an exciting connection between seemingly unrelated phenomena that may serve as a potential explanation for the co-excitation of EUV and Moreton waves, as in the 29 March 2014 event that generated one of the most-clearly observed Moreton waves in addition to strong, concentrated photospheric perturbations. Additionally, our results show that the apparent supersonic nature of Moreton waves may simply be a consequence of a highly inclined acoustic (in reality, magnetoacoustic) wave originating from the interface of the corona with the chromosphere, supporting the original theory proposed by \citep{Uchida}.

\section{Methods}\label{sec:methods}
\subsection{About the Acoustic Model}\label{sec:model}
The acoustic model simulates the propagation of waves throughout the Sun--both the interior and the atmosphere--under the assumption of adiabatic propagation (Equation \ref{eqn:adi}), using the compressible forms of the mass continuity and momentum equations (Equations \ref{eqn:mass} and \ref{eqn:mom}, respectively). The governing equations are given by
\begin{equation}
    \dfrac{DS}{Dt} = \dfrac{D}{Dt}\left(\dfrac{P}{\rho^\gamma}\right)=0 ,
\end{equation}\label{eqn:adi}

\begin{equation}
    \dfrac{D\rho}{Dt}+\rho\nabla\cdot\mathbf{v} =0 ,
\end{equation}\label{eqn:mass}

and
\begin{equation}
    \dfrac{D\mathbf{v}}{Dt} = -\dfrac{1}{\rho}\nabla P + \mathbf{g} + \mathbf{a}_{ext} . 
\end{equation}\label{eqn:mom}
Here, $\rho$ is the mass density, $\mathbf{v}$ is the vector velocity, $P$ is the gas pressure, $\gamma$ is the adiabatic exponent, $\mathbf{g}$ is the acceleration due to gravity, and $\mathbf{a}_{ext}$ is the supplied acceleration. The notation $D/Dt=\partial/\partial t + \mathbf{v}\cdot\nabla$ implies the material derivative. The above governing equations are expanded into a background and perturbed component, and the resulting equations are simplified by removing non-linear perturbations and assuming initial hydrostatic equilibrium.

The background quantities are given in the interior by the Standard Solar Model (SSM) \citep{SSM} and in the atmosphere by the updated VAL-C model \citep{VAL}. The two models are joined where the mass densities are equal; this corresponds to $r = 695.707$ Mm relative to the solar center. The governing equations are solved semi-spectrally, using a fourth-order central finite difference in the radial direction, and spectrally in the $\hat{\theta}$ and $\hat{\phi}$ directions assuming the angular dependence of the perturbed quantities can be separated from the radially-dependent components into spherical harmonics. The 3D simulations are performed by expanding the solution in the spherical harmonics with angular degree up to 6000. The radial dependence of these coefficients is solved along a mesh of 1378 radial grid points, of which 930 correspond to the solar atmosphere with $11.6$ km spacing.

Our initial modeling considered the impact of both electron and proton beams, though the results between the two types do not differ greatly. As such, we show here the results for the proton beam simulation, which ultimately produces a greater sunquake amplitude relative to the Moreton-analogue's amplitude. The proton beam is generated assuming the individual particle energies follow a power-law distribution with spectral index $\delta=5$ and total energy flux $10^{11}$ ergs cm$^{-2}$ s$^{-1}$. For the beam to have finite energy flux, the distribution must be truncated at some cutoff energy, and we choose a modest cutoff energy of $E_c = $~50 keV. Since the acoustic model used in our simulations treats wave propagation as adiabatic, the deposited heat prescribed by the RADYN simulations needs to be included using parameters other than the volumetric energy deposition rate. We, therefore, use two separate methods; in the first method (referred to as the acceleration method), we use the pressure perturbations from the RADYN simulations to derive radial accelerations. In the second method (referred to as the heating method), we supply the acoustic model with pressure perturbations associated with the prescribed heating. In both cases, the heating occurs over just 20 seconds, and the simulation is allowed to evolve for 60 minutes.

\subsection{Computing the Source Functions}

The source functions for the acoustic simulations are derived from the gas pressure perturbations and volumetric energy deposition rates as computed in the RADYN simulations. The incident electron and proton beams are modeled by a power law particle energy flux distribution, given by
\begin{equation}
F_0(E) = \left\lbrace \begin{array}{lc}
    \dfrac{N_0(\delta-1)}{E_c}\left(\dfrac{E}{E_c}\right)^{-\delta} & E\ge E_c \\
     0 & E<E_c
\end{array}\right. ,
\end{equation}
where $E$ is the particle energy, $E_c$ is the energy cut-off (in our simulations, we choose 50 keV for protons and 14 keV for electrons), $\delta$ is the power law index (chosen to be 5), and $N_0$ is the total number density flux (in units of number cm$^{-2}$ s$^{-1}$). The total number density flux is constrained by the total energy flux, such that the integral of the particle energy flux distribution is equal to the total beam energy flux (in our case, $10^{11}$ cm$^{-2}$ s$^{-1}$). The response of the atmosphere to the deposited energy is then computed by RADYN. For the acceleration method, the resulting pressure perturbations are then used to derive the source functions for the acoustic simulation. We consider the acceleration experienced by a fluid element due to the pressure perturbations given by
\begin{equation}
    \mathbf{a}(z,t) = -\dfrac{1}{\rho_0}\dfrac{\partial P^{\prime}}{\partial z} \hat{z},
\end{equation}
where $\mathbf{a}(z,t)$ is the height- and time-dependent acceleration, $rho_0$ is the height-dependent background mass density, and $P^{\prime}$ is the height- and time-dependent gas pressure perturbation. It is assumed that the acceleration is azimuthally symmetric about the beam center, and it decays as a Gaussian with distance from the beam center with a full-width half-maximum of approximately 1.5 Mm. The source function is then supplied to the acoustic model.

For the heating method, we use the energy per unit volume and time deposited by the incident particle beam to derive pressure perturbations for the acoustic model. Since the model is adiabatic, these pressure perturbations are necessary as there is no heat source term in our entropy equation. The corresponding pressure perturbation, assuming an ideal gas, is given by
\begin{equation}
\dfrac{\partial P^\prime}{\partial t}(z,t) = Q^{\prime}_{\text{vol}}(\gamma-1),
\end{equation}
where $P^{\prime}$ is the pressure perturbation, $Q^{\prime}_{\text{vol}}$ is the volumetric heating rate in ergs cm$^{-3}$ s$^{-1}$, and $\gamma$ is the heat capacity ratio. This equation is included as a source term in the expanded form of Equation \ref{eqn:adi}, and the heating is assumed to have the same horizontal dependence as the acceleration method.

\section{Results}\label{sec:results}

\begin{figure}[h!]
    \centering
    \includegraphics[width=\textwidth]{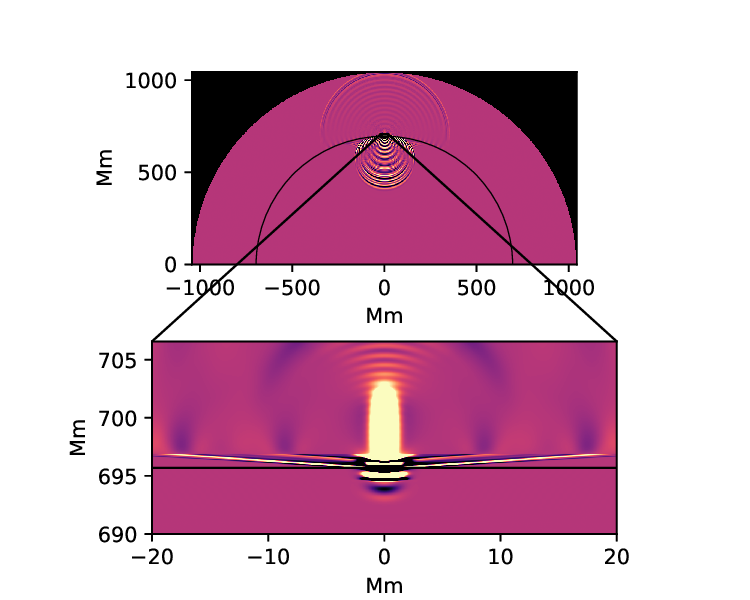}
    \caption{A cut-through of the acceleration method simulation at t=2500 seconds showing the scaled radial velocity. The inset figure shows the same simulation at an earlier time (t=200 seconds), which demonstrates the scale at which the local results in this work are presented. An animation of this figure is available.}\label{fig:full_sun}
\end{figure}

\begin{figure}[h!]
    \centering
    \adjincludegraphics[width=\textwidth,trim={{0.1\width} 0 {0.1\width} 0}]{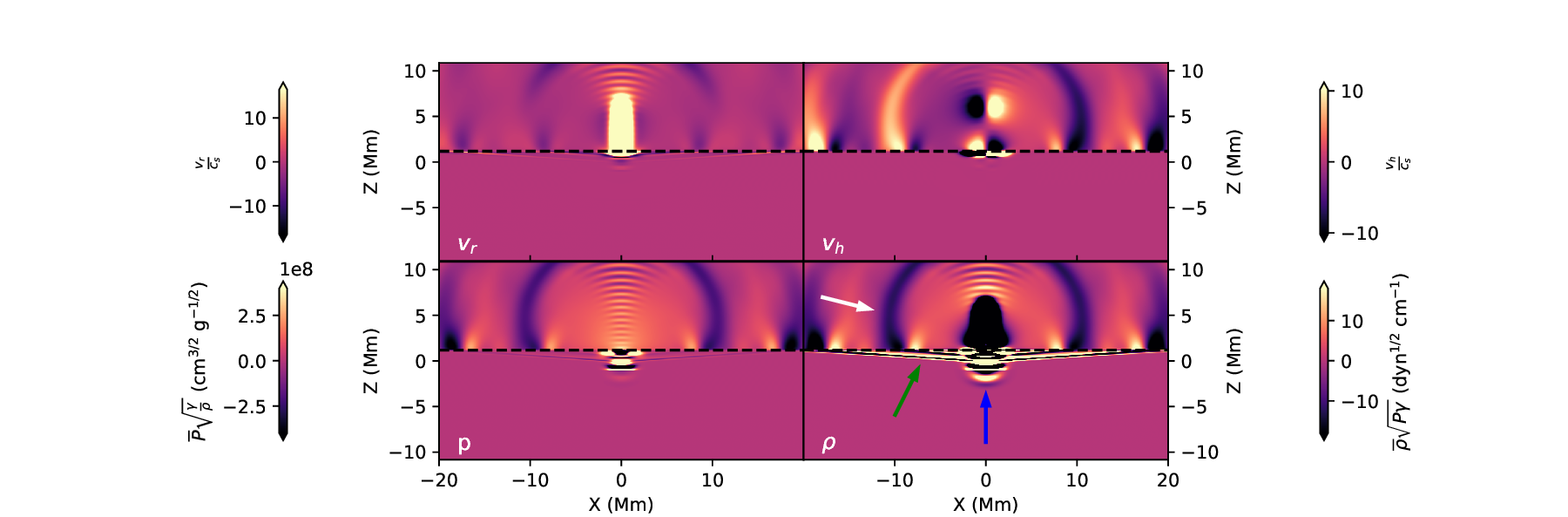}
    \caption{A cut-through of the proton beam acceleration method simulation at t=200 seconds showing the radial velocity (top-left), $\theta$-component of velocity (top-right), pressure perturbation (bottom-left), and density perturbation (bottom-right). The light and dark colors indicate positive and negative quantities, respectively. The simulation reveals the propagation of a coronal wave that intersects the transition region (dashed line) to form a strong chromospheric wave that propagates mainly downward. The white arrow marks the coronal wave, the green arrow marks the chromospheric portion of the wave after refracting across the transition region, and the blue arrow marks the sunquake wavefront as it begins to form. An animation of this figure is available.}\label{fig:acc_still}
\end{figure}

We begin with the results from the proton beam treated with the acceleration method that produces globally propagating waves in three distinct layers: the corona, the chromosphere, and the solar interior. These waves are analogues of those observed on the Sun: LCPFs, Moreton waves, and sunquakes, respectively. The coronal and internal waves are clearly visible in Figure \ref{fig:full_sun}, where the top boundary has been extended to 1.5 $R_\odot$ to exaggerate the coronal wavefront. Beyond $z\approx 10$~Mm, the atmosphere is stratified such that the sound speed is constant while the density and pressure decrease with gravitational stratification. The coronal wave propagates the fastest until the internal wave (sunquake) penetrates more deeply, where the sound speed is greater. The Moreton wave analogue, however, propagates through the chromosphere, occupying a much smaller region that is difficult to see in the global picture of Figure \ref{fig:full_sun}. We, therefore, examine the generation and propagation of the Moreton-analogue wave on a more local scale so that fine features are more visible, as shown in Figure \ref{fig:acc_still}. The quantities here have been scaled so that the wavefront is visible across the wide range of vertical positions where the density and gas pressure change by several orders of magnitude. Both components of velocity are divided by the local sound speed, the pressure perturbation is scaled by $c_s/\sqrt{P_0}$, and the density perturbation is scaled by $c_s/\sqrt{\rho_0}$. Here, $c_s=\sqrt{P\gamma/\rho}$ is the background adiabatic sound speed. The dashed lines in the panels of Figure \ref{fig:acc_still} mark the location of the transition region, which in the RADYN simulations begins at $z=1.175$ Mm. The modeled transition region is thin, spanning only 20 km, where the plasma density drops two orders of magnitude from the upper chromosphere to the low corona.

At the beginning of the acoustic simulation, the heating from the proton beam causes strong chromospheric evaporation seen in the top-left panel of Figure \ref{fig:acc_still} as an over-saturated bright feature (upflow) at the beam core. The generation of a coronal wave follows shortly after the simulation initiates, and the upflows from the evaporation dominate the wavefront in the upper portion of the simulation. However, this upward velocity is not the source of the coronal wavefront; instead, there is a weaker, downward-velocity wavefront that originates near the transition region. This wavefront forms between 10 and 20 seconds after the acoustic simulation begins, and is formed by the following chain of preceding events. The upflows prescribed by the acceleration method form an expanding region of over-pressure within the first few seconds, which drives flows away from the beam core. These outflows collide with the surrounding stationary plasma, increasing the local density. The over-dense plasma then falls since hydrostatic equilibrium is lost, and this forms the downward-velocity wavefront near the transition region.

\begin{figure}[h!]
    \centering
    \includegraphics[width=0.5\textwidth]{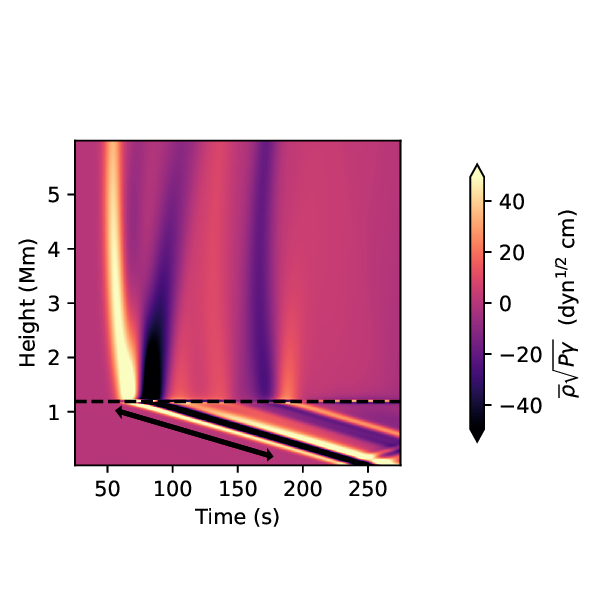}
    \caption{Time-height diagrams of the scaled density perturbation at a point 20 Mm from the beam core, showing the vertical propagation of the initial wavefront in the simulation treated with the acceleration method.}\label{fig:tz}
\end{figure}

Since this wavefront forms closest to the transition region, it is also the first to enter the chromosphere, with alternating upflows and downflows following behind it. The coronal portion of the wavefront expands evenly in all directions, and as it propagates horizontally along the upper transition region, the wave is refracted and begins traveling downwards into the chromosphere. There is a significant difference between the sound speeds just above and just below the transition region (75 km s$^{-1}$ and 7 km s$^{-1}$, respectively), and the refracted wave travels almost entirely vertically within the chromosphere. In essence, each point along the transition region acts like a wave generator, exciting a wave that travels downward in the chromosphere at the local sound speed. The time delay between each successive point is given by the travel time for the coronal portion of the wave, which is necessarily shorter than a wave propagating horizontally in the chromosphere due to the greater sound speed. The left panel of Figure \ref{fig:tz} shows the vertical propagation of the wavefront in the acceleration method simulation at a point 20 Mm away from the beam core. The wavefront arrives at all heights in the low corona almost simultaneously but is slightly delayed closer to the transition region, where the sound speed is smaller. The wave then propagates downwards in the chromosphere, with the black arrow tracing out the path taken by a wave moving at the local sound speed.

\begin{figure}[h!]
    \centering
    \includegraphics[width=0.5\textwidth]{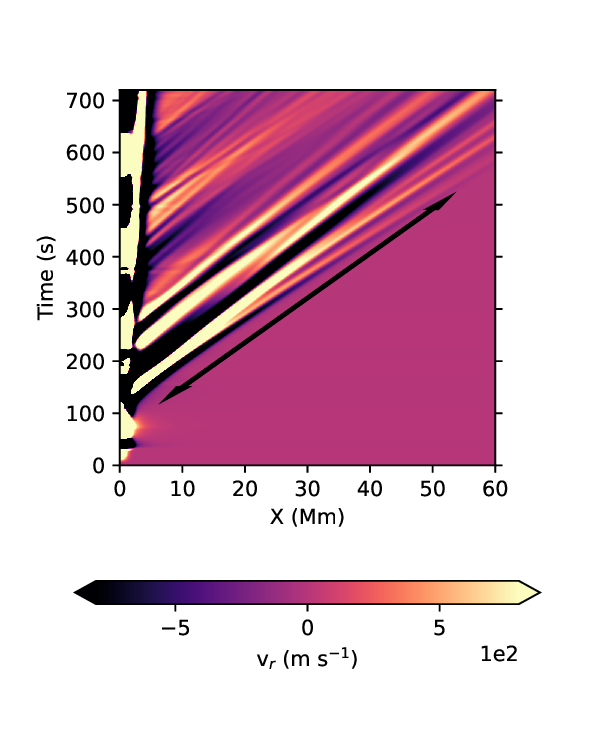}
    \caption{Time-distance diagram of the radial velocity at z=500 km (the mid-chromosphere). The black arrow denotes the apparent mean propagation speed (117 km s$^{-1}$) of the Moreton-wave analogue.}\label{fig:proton_both_td}
\end{figure}

The effect of this chain of wave generators, when viewed along the horizontal plane, is what appears to be a supersonic wavefront traveling horizontally through the chromosphere. In the left panel of Figure \ref{fig:proton_both_td}, we show the time-distance diagram of the acceleration method simulation at $z=500$ km, where the radial velocity is averaged over successively larger great circle distances. This provides a useful way for studying wave behavior, as in a similar way to the time-height diagrams in Figure \ref{fig:tz}, the speed of the wave can be found from its slope in the diagram. When viewed in this manner, the chromospheric wavefront appears to be traveling at an average speed of 117 km s$^{-1}$ (the black arrows in Figure \ref{fig:proton_both_td}), significantly faster than the local sound speed of 7 km s$^{-1}$. However, both Figures \ref{fig:acc_still} and \ref{fig:tz} show that the wavefront is propagating at the local sound speed, and only appears supersonic because the source---the coronal wavefront---is traveling this quickly. Additionally, the chromospheric wavefront appears to speed up slightly, from 104 km s$^{-1}$ at $x=20$ Mm to 130 km s$^{-1}$ at $x=50$ Mm; this occurs because the portion of the coronal wavefront that reaches greater distances has propagated partially at higher heights where the sound speed is greater.

\begin{figure}[h!]
    \centering
    \includegraphics[width=\textwidth]{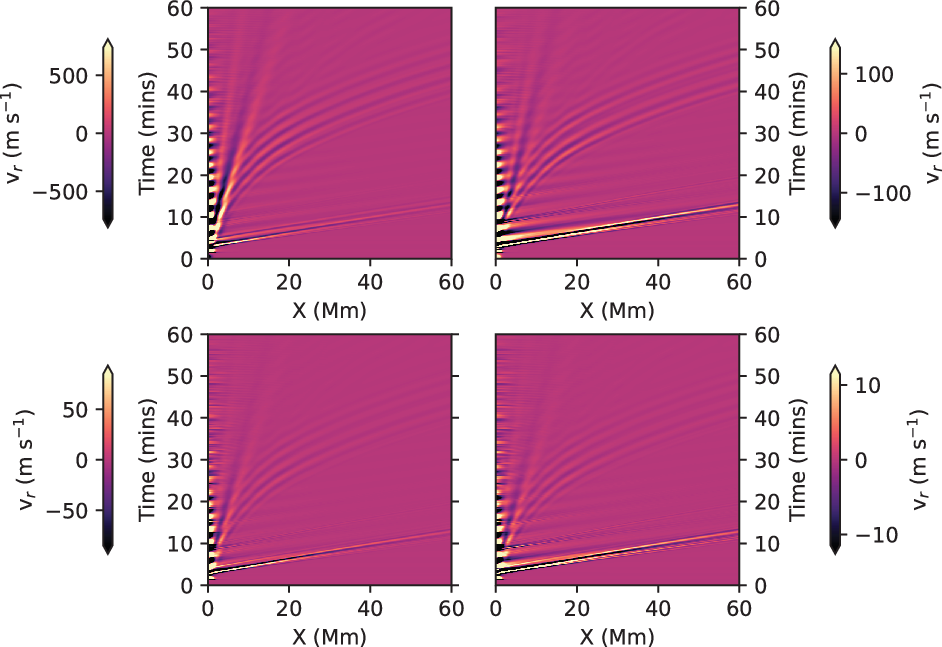}
    \caption{Time-distance diagrams at $z=100$ km for the proton beam simulations (left column) and electron beam simulations (right column), using the acceleration method (top row) and heating method (bottom row). The Moreton-analogue wave appears around $t=5$ minutes and is strongly inclined, while the sunquake appears later with a gradually increasing inclination. Note, the time and length scales are greater than those for Figure \ref{fig:proton_both_td}.}\label{fig:sq_td}
\end{figure}

There are only two qualitative differences between the two methods used to insert the effects of the proton beam into the acoustic simulation. First, the wavefront amplitudes in the heating method simulation are smaller than those in the acceleration method by about a factor of two. And second, it takes slightly longer for the effects of the beam in the heating method to appear beneath the beam core. This second difference occurs because the acceleration method prescribes motion (albeit weak) in the deeper portions of the chromosphere, whereas the pressure perturbations prescribed by the heating method are limited mostly to the transition region and the highest portions of the chromosphere. This time delay between beam initiation and the appearance of its effects in the lower portions of the atmosphere is more apparent in Figure \ref{fig:sq_td}, which shows time-distance diagrams at $z=100$ km for both the proton (left column) and electron (right column) beams modeled by the acceleration (top row) and heating (bottom row) methods. Both the proton and electron beam simulations treated with the acceleration method show an essentially instantaneous, though weak, response at the photosphere. Additionally, the initial radial velocity between the two beam types (proton vs electron) is of opposite sign. While the explanation for this isn't immediately clear, the proton beams deliver significantly more momentum flux than the electron beams due to their higher mass per particle. The downward momentum flux delivered by the proton beam is perhaps sufficient to overcome the upflows driven by chromospheric evaporation such that there is an initial net downward motion at the photosphere. The momentum flux delivered by the electron beam, meanwhile, may not be enough to overcome the evaporation-driven upflows.

An additional difference between all four simulation cases is in the radial velocity amplitudes of the produced Moreton-analogue wave and sunquake at the photosphere. In general, the acceleration method produces waves with an amplitude about an order of magnitude greater than the heating method for both the proton and electron beams. In line with the greater amount of momentum flux deposited by the proton beams, the radial velocity of both the sunquake and Moreton-analogue waves in the electron beam simulations is consistently less than the corresponding proton beam simulation. Furthermore, the ratio between the Moreton-analogue wave's and the sunquake's amplitude is an interesting quantity to consider as it provides some hint at potential observability. The sunquake amplitude away from the beam source is maximized around $x=18$ Mm due to focusing by the subsurface sound speed gradient, so we compare the relative amplitudes here. For the proton beams, this ratio is $2.3$ and $10.1$ using the acceleration and heating methods, respectively. For the electron beams, the ratio is $12.3$ and $21.1$ using the acceleration and heating methods, respectively. The electron beams consistently produce a stronger Moreton-wave analogue relative to the generated sunquake; this, coupled with the overall low sunquake amplitude, indicates that electron beams may be primarily responsible for Moreton waves observed without a corresponding eruption.

\section{Discussion and Conclusion}\label{sec:diss}

In this work, we have shown that particle beams are capable of simultaneously generating coronal and Moreton-like waves as well as sunquakes, and we show how the refraction of coronal waves across the transition region explains the apparent supersonic nature of Moreton waves. We have, however, made several simplifying assumptions. First, true coronal waves and Moreton waves are likely to be generated by shocks associated with beam heating, the propagation of which is certainly not well-described by the linear, adiabatic nature of our acoustic simulations. This deficiency is especially clear in Figure \ref{fig:acc_still} where the upflows at the beam core and atmospheric wavefronts have supersonic amplitudes. Our goal, though, is to demonstrate the qualitative behavior of the wave dynamics resulting from a more realistic excitation source of sunquakes. A more exact modeling of these waves requires non-linear MHD simulations to capture the generation of shocks and the interaction of the waves with a background magnetic field.

Such a background magnetic field has not been included in our simulations, though we attempt to address this---at least partially---through our use of two different methods for inserting the effects of beam heating into the acoustic simulation. Recent 2.5D flare models have shown how the initial background magnetic field strength affects the perturbation of the chromosphere during beam heating; these models have shown that as the background field strength increases, the chromospheric evaporation penetrates more deeply, occurs over shorter time scales, and produces faster downflows \citep{2.5Dflare}. The heating model used in our acoustic simulations is similar to the weak background magnetic field case, where only the upper chromosphere is significantly perturbed by the beam heating. Meanwhile, the acceleration model is more consistent with the strong background magnetic field case, where the beam heating affects nearly the entire chromosphere and initiates noticeably stronger flows. 

The background magnetic field is expected to play an important role not only in the excitation of Moreton and coronal waves but also in their propagation. First, the presence of a magnetic field allows these waves to travel more quickly, at the fast magnetosonic wave speed, as opposed to simply the sound speed. While we cannot address this aspect directly through our simulations, the fast magnetosonic wave speed likely changes with height in a similar fashion to the sound speed in the presence of a uniform magnetic field. In this way, the stratification of wave speed in our simulated atmosphere is qualitatively similar to the true wave speed stratification. Additionally, a curious aspect of Moreton and coronal waves is their tendency to induce oscillations in filaments and prominences after having passed through these structures \citep{Cabezas}. Unfortunately, we are not able to study such effects in our simulations.

Finally, our acoustic simulations do not include radiation, so we are unable to model how the simulated waves would appear in observations---for example, in 304 \AA~ or in H$\alpha$. Radiation is also expected to play an important role in wave damping, though we address this aspect by artificially damping waves according to observed damping time scales. These damping time scales are obtained from the power spectra of quiet-Sun pressure waves (p-modes), and the full procedure is discussed in more detail in \citep{SQ_model}. This wave-damping scheme is somewhat insufficient, however, as it was designed with realistic sunquake damping in mind, so the scheme is better suited for damping internal waves than it is for atmospheric waves. This is evident in the reflection of the coronal wave at the upper boundary in the acceleration method simulation despite the use of non-reflecting boundary conditions and in the reflection of the Moreton-like wave at the photosphere which causes additional coronal waves to be excited. However, multiple coronal wavepackets have sometimes been found in observations and in other simulations \citep{Devi,Chen}. The imbalance in wave damping between the sunquake and Moreton-analogue wave may also skew the ratio of their amplitudes, such that the amplitude of the Moreton-analogue at the photosphere presented in Figure \ref{fig:sq_td} may be significantly weaker in reality. As such, it may be the case that proton beams are incapable of producing an observable Moreton wave, or perhaps one at all. This would further support the notion that proton beams are responsible for the generation of sunquakes, a hypothesis that is currently being explored \citep{Sadykov2023}.

Despite our simplifying assumptions, we are also able to reproduce another important feature of Moreton waves aside from their apparent supersonic propagation speed. Observations of Moreton waves in H$\alpha$ often show absorption in the red wing and enhancement in the blue wing of the line \citep{Cabezas}, and Doppler analysis of H$\alpha$ and other lines indicate that the leading edge of a Moreton wave is a downflow or relaxation front \citep{Doppler}. In both the acceleration method and heating method simulations, we observed downflows at the leading edge of the Moreton-like wave. Tracking the formation of this wavefront reveals that this initial downflow is likely fundamentally linked to chromospheric evaporation, as it originates from the overdensity outside of the evaporation region. This is further supported by the formation of downflows at the leading edge of the Moreton-analogue wave in both methods of simulation, despite their different treatments.

To summarize our results, we find evidence for the generation of coronal disturbances---analogous to EUV waves or LCPFs---in our acoustic simulations of sunquake generation. These disturbances propagate from the corona and into the chromosphere, where they appear as supersonically-traveling wavefronts analogous to Moreton waves. This apparent supersonic propagation in the chromosphere is merely a consequence of the corona's faster sound speed. The generation of large-scale coronal waves by proton or electron beams can also help explain Moreton and EUV waves observed without an accompanying CME, particularly for the electron beams which have high-amplitude EUV and Moreton waves but an essentially undetectable sunquake because of the low amplitude relative to background convective motions. Additionally, we find that the initial radial velocity perturbation in the photosphere beneath the sunquake source has a different sign depending on the particle beam content (proton vs electron). While this difference may be a by-product of our simulation setup, it may also be a means of characterizing observed sunquake sources.

\begin{acknowledgements}
    This work was supported by NASA grants 80NSSC23K0097, 80NSSC19K1436, 80NSSC20K0302, and 80NSSC19K0268, as well as NSF grant 1916509.
\end{acknowledgements}

\bibliography{SQ_atmo_bib}{}
\bibliographystyle{aasjournalv7}



\end{document}